# Nanocircuit transmitting microwave and terahertz harmonics generated with a mode-locked laser—revised

Mark J. Hagmann,[1,2] and Isaac Martin[2]
[1]Department of Electrical and Computer Engineering, University of Utah, Salt Lake City, Utah 84112
[2]NewPath Research L.L.C., 2880 S. Main Street, Suite 214, Salt Lake City, Utah 84115.

**Abstract**

Earlier we focused a mode-locked laser on the junction of a scanning tunneling microscope (STM). This superimposed currents at the first 200 harmonics of the laser pulse-repetition rate on the DC tunneling current. The power at each harmonic is inversely proportional to the square of its frequency. However, other measurements suggest that within the tunneling junction the harmonics only decay as the exponent of minus the square of the product of the harmonic number and the laser duty cycle which would not be noticeable below 45 THz. Now we apply this effect in a nanocircuit to mitigate the roll-off in power as the square of the frequency. This nanocircuit has an optical antenna to receive the laser radiation, symmetric field emission diodes to generate the harmonics, and filters to select microwave or terahertz harmonics that are transmitted by a second antenna. In this unique approach, when transmitting over a chosen bandwidth $\Delta f$ that is proportional to $f_0$, the frequency at the center of this bandwidth, the output power is proportional to the square of the center frequency because of the fixed spacing of the harmonics.

**Introduction**

This is a revision of the recent arXiv-2108-13832 (2021) where the analysis is unchanged but the format is modified and other references are added to increase the clarity of the presentation.

Section 1 cites publications describing our applications of quantum analysis that led to discovering the microwave frequency comb (MFC) with an STM. However, this paper describes simpler methods to design the nanocircuits that are based on the analysis of the harmonics measured with the STM. Most of the derivations for this analysis were moved to the appendices. Note that, while individual harmonics can be transmitted by the nanocircuit at microwave frequencies, it would not be possible to separate them at terahertz frequencies with conventional filters so instead we will transmit bands of adjacent harmonics which increases the output power. Section 1 introduces the presence of a cutoff that limits the harmonics to a maximum frequency of 45 THz, which is further considered in Section 4.

Section 2 describes the new nanocircuit, including a sketch.

Section 3 describes how the incident laser radiation is coupled to each field emission diode (FED) in the nanocircuit.

Section 4 contains the derivation of a general expression for the current in each FED which includes the effect of the cutoff.

Section 5 derives a general expression for the total current below, at, or above the cutoff.

Section 6 derives an expression for the total current in sets of harmonics that are well below the cutoff frequency. It also provides an expression for the number of nulls in the total current by setting the first term in Eq. (15) to zero. Appendix 4 shows that setting the second term in that equation to zero gives the same number of nulls.

Section 7 shows that the total current in the harmonics would be infinite in an ideal system having no cutoff.

Section 8 contains a derivation for the total RMS current and the total power in a set of adjacent harmonics that is well below the cutoff.

Section 9 considers possibilities for the antenna that transmits the harmonics.

Section 10 presents seven simulations of the current as a function of time during a single period of the laser for different sets of adjacent harmonics. Figure 13 shows the limiting case as the number of harmonics is infinite while neglecting the effects of the cutoff.

Section 11 presents suggestions for optimizing the nanocircuits in microwave and terahertz applications, and Section 12 is the summary and conclusions.

Appendix 1 derives a closed-form expression for the total current in any set of adjacent harmonics that is well below the cutoff.

Appendix 2 derives a closed-form expression for the total current in any set of adjacent harmonics that is near the cutoff.

Appendix 3 derives a general closed-form expression for the total current in any set of adjacent harmonics that may include those below, near, or above the cutoff.

Appendix 4, which is referred to in Section 6, completes the presentation in that section by showing that setting the second term in Eq. (15) to zero gives the same values of $\omega_0 t$ for the nulls in the total current as those that were obtained in Section 6 by setting the first term in Eq. (15) to zero.

## 1. Background

First, we used quantum theory to simulate laser-assisted tunneling and determined that it was possible to use this mechanism to generate microwave radiation by either laser-assisted tunneling or laser-assisted field emission where the tunneling is followed by classical transport [1], [2], [3], [4].

Later, guided by simulations, we focused a mode-locked Ti:Sapphire laser on the tunneling junction of a scanning tunneling microscope (STM) to generate a unique MFC [5]. There are hundreds of harmonics at integer multiples of the 74.25 MHz laser pulse-repetition frequency. This may be understood because the tip-sample distance is much shorter than the wavelengths of the laser so at nanoscale the effectively the laser superimposes a time-dependent voltage on the DC bias of the STM. Thus, at present our emphasis is not on quantum theory, but rather on determining the properties of the measured harmonics and designing devices that are based on these observations.

The 200[th] harmonic, at 14.85 GHz, has a power of only 2.8 atto-watts with an applied DC tunneling current of 10 μA. However, its signal-to-noise ratio is 25 dB because the FWHM (full-width at half-maximum) linewidth of 1.2 Hz for each harmonic reduces the received background noise. However, this measured linewidth is an upper bound as a convolution of the actual linewidth with the response of the spectrum analyzer. Thus, the quality factor (Q) exceeds $10^{10}$ as the state-of-the-art for low-noise in microwave measurements [6].

Figure 1 is a block diagram for the system we used to measure the microwave harmonics with a miniature 50-Ω coaxial cable connecting the STM tip to a digital spectrum analyzer.

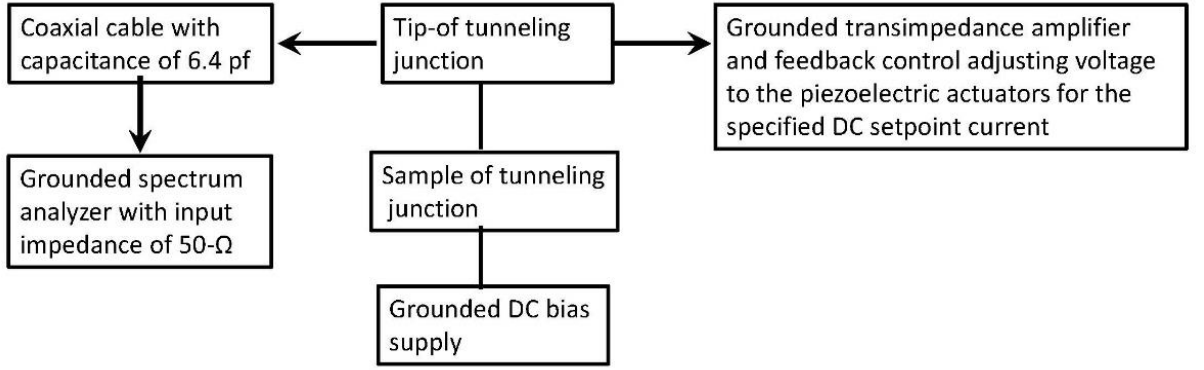

Fig. 1. Block diagram of the system used to measure the microwave harmonics with an STM.

The resistance of the tunneling junction in the STM is much greater than the characteristic impedance of the cable. Thus, the cable acts as a capacitance of 6.4 pF across the 50-Ω input impedance of the spectrum analyzer. The current at each harmonic is divided between the spectrum analyzer and the capacitive shunt presented by the cable. This forms a low-pass filter, with a 3dB point of 500 MHz. Thus, the total power measured by the spectrum analyzer for the first N harmonics is given by Eq. (1) where R is the input impedance of the spectrum analyzer, C is the shunting capacitance of the miniature coaxial cable, and $\omega_I$ is the angular frequency of the I-th harmonic. Later in this paper we will consider the fundamental limit to this type of measurement that becomes dominant at a frequency of 45 THz [7]. If it were not for this effect there would be no conservation of energy.

For typical parameters that are used in the STM measurements the second term in the denominator of Eq. (1) is dominant so Eq. (2) may be used. Thus far we have only measured the first 200 harmonics with the STM, and the roll-off in the power at these harmonics suggests that all of the $P_I$ are equal. Thus, we set $P_I$ to $P_1$, the power at the first harmonic. We define $\omega_I$ as $I\omega_0$ where $\omega_0$ is the angular frequency corresponding to the pulse-repetition rate of the laser to obtain Eq. (3). In the limit as N becomes large Eq. (3) approaches Eq. (4) for an estimate of the total power of the harmonics in the STM measurements, where now $f_0$ is the pulse-repetition rate of the mode-locked laser.

$$P_{tot} = \sum_{I=1}^{N} \frac{P_I}{1+\omega_I^2 R^2 C^2} \quad (1)$$

$$P_{tot} = \frac{1}{R^2 C^2} \sum_{I=1}^{N} \frac{P_I}{\omega_I^2} \quad (2)$$

$$P_{tot} = \frac{P_1}{\omega_0^2 R^2 C^2} \sum_{I=1}^{N} \frac{1}{I^2} \quad (3)$$

$$P_{tot} = \frac{P_1}{24 f_0^2 R^2 C^2} \quad (4)$$

A power of -86 dBm (2.5 pW) was measured in the STM at the first harmonic where $f_0$ is 74.25 MHz when using an R of 50 Ω and C equal to 6.4 pF with a DC tunneling current of 50 μA [8]. Thus, Eq. (4) shows that the total power for all of the harmonics was 73.8 times greater than that at the first harmonic, or 73.8 x 2.5x10$^{-12}$ W which is 0.18 nW.

We consider the possibility that these harmonics are not merely a perturbation of the DC tunneling current due to the nonlinear dependence of the current on the applied bias voltage in an STM. This becomes clear in our analysis and simulations of Scanning Frequency Comb Microscopy (SFCM) where the harmonics would be used for feedback control of the tip-sample distance in an STM without an applied DC bias.

We acknowledge that SFCM would require resistive samples such as semiconductors, and while it has been analyzed, and simulated, it has not yet been tested [9],[10],[11]. In SFCM, first tunneling would be established using feedback control of the tip-sample separation to provide a specified setpoint value for the DC tunneling current. Then the DC bias would be turned off and the feedback control of the tip-sample distance would be based on the power that is measured in the MFC. Feedback control of the tip-sample distance would be accomplished by adjusting the tip-sample distance to maximize the power at the harmonics. This may be understood because, at the distance for maximum power, (1) reducing the tip-sample distance increases the spreading resistance by causing the harmonics to pass through a cross-sectional smaller area in the sample, while (2) increasing the tip-sample distance increases the tunneling resistance which also reduces the current in the harmonics.

Now we are studying the possibility of fabricating nanocircuits in which much higher harmonics could be generated without the need for feedback control while mitigating the roll-off in the output power at the higher harmonics with an STM. The nanocircuits will use symmetric field emission diodes that are much smaller than the tip and the sample electrodes in an STM. This increases the ratio of the surface area to the volume to provide greater heat transfer. This facilitates the use of much greater current densities at the harmonics. Optical antennas will receive the radiation from a mode-locked laser and an antenna will transmit selected harmonics at microwave and/or terahertz frequencies (patent-pending).

## 2. Description of the nanocircuit

Figure 2 is a sketch of the design for the new nanocircuit. Here "FED" denotes each of the two symmetrical field emission diodes that replace the STM tunneling junction. No semiconductors are required. Each FED will have a small gap in the metal lead, coated with a refractory metal such as iridium, and ion beam lithography will be used to form a sub-nanometer gap in the refractory metal. Each FED is symmetric in order to have the same response with either polarity. All of the components are attached to a substrate having a low dielectric constant to limit coupling to the transmitting antenna and other components. This sketch is not made to scale.

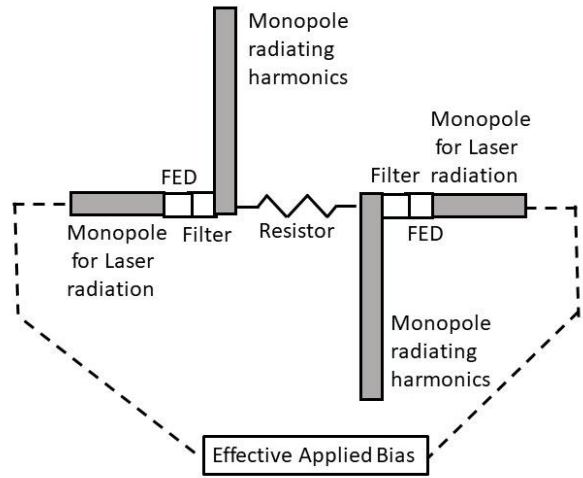

Fig. 2. Sketch of the nanocircuit generating microwave and terahertz harmonics.

Instead of focusing a laser on the tunneling junction of an STM, the laser will excite the monopole optical nanoantennas shown at the left and right ends of the sketch. This will cause a voltage having the full spectrum of the laser across each FED that is labeled as the "effective applied bias" to relate this to our STM measurements. The mode-locked laser will have a precision power splitter with a phase shifter to be focused simultaneously on the two optical nanoantennas.

The two filters prevent the spectrum of the laser from reaching the circuit between them. Thus, only the current at the harmonics generated in the two field emission diodes passes through the resistor at the center of the sketch. This resistor is across the feed-point for the dipole that is formed by the two vertical monopole antennas. Thus, the voltage across this resistor feeds the vertical dipole to radiate at microwave or terahertz harmonics that are selected by the two filters. Nanoscale resistive L-type networks may be included to provide broadband impedance matching.

## 3. Method for generating the current in the field emission diodes

Focusing a laser normal to the outer electrode of each FED would not cause optimum electron transport within the FED because the incident electric field from the laser would be tangential to the electrode. One method to solve this problem is shown in Fig. 3 where the laser radiation passes horizontally across the surface of the outer electrode in each FED. The electric field from the laser is normal to the surface of this electrode to cause the electrons to be displaced vertically within the metal as shown to create electrical charges with opposite signs on the upper and lower surfaces. Thus, there is a vertical electric field between the two electrodes of each FED to cause electron transport between them as shown in Fig. 3. In this case the optical antennas receiving the laser radiation could be vertical metallic nanotips attached to the electrode. It is also possible for the surface of the electrode to act as an antenna. The width of each electrode for the field emission diodes is much less than the wavelengths of the laser so this is a quasistatic process. This sketch is not made to scale.

A FED is shown in Fig. 3 because it is simpler to fabricate than a tunneling diode (TD) in this application. Direct tunneling is only possible with an extremely small gap between the two electrodes which must be maintained by either feedback control as in an STM or a semiconductor spacer. However, feedback control would be difficult in this application and a semiconductor would breakdown if the laser is too intense. Field emission would be satisfactory in this application if the length for classical propagation that follows tunneling is small enough that the delay it causes would not interfere with generating the harmonics.

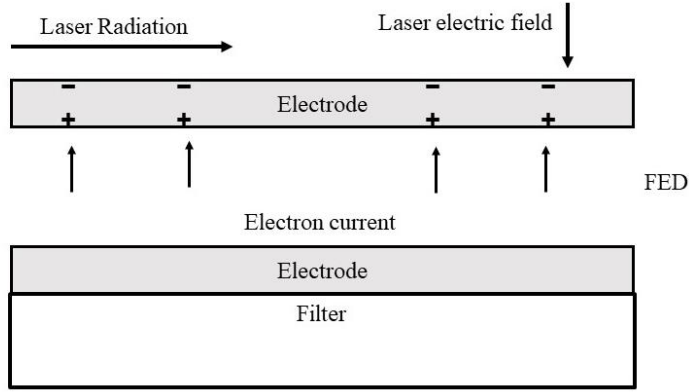

Fig. 3. Sketch of the field emission diode with the electric field of laser normal to the outer electrode.

## 4. General expression for the current in each field emission diode

In a previous analysis we showed that, if the harmonics could be measured within the tunneling junction of an STM, the power at the k-th harmonic would be given by Eq. (5) where $P_1$ is the value at the first harmonic [7]. Thus, Eq. (6) gives the current at the k-th harmonic where $I_1$ is the magnitude oof the current at the first harmonic. Here "T" is the period for each cycle of the mode-locked laser and τ is the duration of each laser pulse. The duty cycle, τ/T, is extremely small. Thus, in STM measurements [5], where a power of -120 dBm was measured at the first harmonic with a spectrum analyzer having an input impedance of 50 Ω, and $I_1$ was approximately 6 nA.

$$P_k = P_1 e^{-\pi^2 (k-1)^2 \left(\frac{\tau}{T}\right)^2} \tag{5}$$

$$I_k = I_1 \sin(k\omega_0 t) e^{-\frac{\pi^2 \tau^2}{2T^2}(k-1)^2} \tag{6}$$

Equation (6) is simplified to give Eq. (7) where A is an extremely small parameter defined in Eq. (8) and k is the harmonic number. We refer to the phenomena occurring where $Ak^2$ is comparable to unity as the "cutoff" because the exponential decay of the current that is caused by the physics is first prominent in the harmonics above that point.

$$I_k = I_1 \sin(k\omega_0 t) e^{-Ak^2} \tag{7}$$

$$A \equiv \frac{\pi^2}{2}\left(\frac{\tau}{T}\right)^2 \tag{8}$$

The analysis that resulted in Eq. (5) did not allow for the effects of the capacitance C shunting the FED or the load resistance R, but these effects were included in Eqs. (1), (2), (3), and (4) for the analysis with an STM. However, the capacitance of each FED in the nanocircuit may be much less than 1 femto-Farad so we make the approximation of neglecting these effects. Thus, with the nanocircuits, we add the powers at all of the harmonics to obtain the total power to the load.

## 5. Use of several microwave harmonics or many harmonics at terahertz frequencies

In applying this technology, the part of the full spectrum of harmonics that is transmitted is determined by the two filters as well as the properties of the antenna transmitting the harmonics. At microwave frequencies it is possible to transmit one or several individual harmonics. However, at terahertz frequencies the spacing between adjacent harmonics is much smaller than their individual frequencies so present filter technology cannot be used to separate the individual harmonics. Sets of adjacent harmonics are of special interest because the larger number of harmonics used at terahertz frequencies (e.g., 1000) can transmit much higher power.

Here we consider the effect of transmitting a set of adjacent harmonics where the fundamental frequency $\omega_0$, which is the angular form of the pulse repetition rate for the laser, is defined in Eq. (9). Thus, at time t, the total current from these harmonics is given by Eq. (10).

$$\omega_k \equiv k\omega_0, \text{ where the integer } k_1 \leq k \leq k_2 \quad (9)$$

$$I(t) = I_1 \sum_{k=k_1}^{k_2} \sin(k\omega_0 t) e^{-Ak^2} \quad (10)$$

Eq. (10) is solved to obtain closed-form algebraic expressions for the total current in sets of adjacent harmonics that are well below the cutoff in Appendix 1 and near the cutoff in Appendix 2, as well as for the general case in Appendix 3.

## 6. Superposition of sets of adjacent harmonics that are well below cutoff

The current at each harmonic is zero at the exact start and end of each period of the laser. If the laser had an infinite coherence length and operated well below the cutoff, the total current during each period would have maxima and minima at the ends which decay progressively to smaller magnitudes near the center of the period. This is shown in the examples for Section 14. Now, we examine the mechanism for these effects.

Regrouping the terms in Eq. (A1-3) in Appendix 1 gives Eq. (11). Then, using the trigonometric identities in Eqs. (12) and (13), we obtain Eq. (14). Finally, simplifying Eq. (14) gives Eq. (15).

$$I(t) = \frac{I_1}{2} \left[ \begin{array}{l} \sin(k_1\omega_0 t) + \sin(k_2\omega_0 t) \\ + \cot\left(\frac{\omega_0 t}{2}\right)\left[\cos(k_1\omega_0 t) - \cos(k_2\omega_0 t)\right] \end{array} \right] \quad (11)$$

$$\sin(\alpha) + \sin(\beta) \equiv 2\sin\left(\frac{\alpha+\beta}{2}\right)\cos\left(\frac{\alpha-\beta}{2}\right) \quad (12)$$

$$\cos(\alpha) - \cos(\beta) \equiv -2\sin\left(\frac{\alpha+\beta}{2}\right)\sin\left(\frac{\alpha-\beta}{2}\right) \quad (13)$$

$$I(t) = \frac{I_1}{2} \left[ \begin{array}{l} 2\sin\left[\frac{(k_2+k_1)\omega_0 t}{2}\right]\cos\left[\frac{(k_2-k_1)\omega_0 t}{2}\right] \\ +2\sin\left[\frac{(k_2+k_1)\omega_0 t}{2}\right]\sin\left[\frac{(k_2-k_1)\omega_0 t}{2}\right]\cot\left(\frac{\omega_0 t}{2}\right) \end{array} \right] \quad (14)$$

$$I(t) = I_1 \sin\left[(k_2+k_1)\frac{\omega_0 t}{2}\right] \left[\begin{array}{l} \cos\left[(k_2-k_1)\frac{\omega_0 t}{2}\right] \\ + \sin\left[(k_2-k_1)\frac{\omega_0 t}{2}\right]\cot\left(\frac{\omega_0 t}{2}\right) \end{array}\right] \quad (15)$$

The first term in brackets for Eq. (15) shows that the total current is zero at a sequence of times given by Eq. (16) where n may be any positive integer less than $k_1 + k_2$. Note that these nulls necessarily occur at closer intervals when either $k_1$ or $k_2$ is increased. For example, with $k_1$ equal to 1 and $k_2$ equal to 5 there would be 5 nulls within each period of the laser, where $t_n/T$ is 1/6, 2/6, 3/6, 4/6, and 5/6. Also, with $k_1$ equal to 1 and $k_2$ equal to 10 there would be 10 nulls. In general, there are $k_2 - k_1 + 1$ nulls which is the number of harmonics.

$$\frac{t_n}{T} = \frac{n}{(k_1 + k_2)} \quad (16)$$

An analysis in Appendix 4 shows that the zeros obtained by setting the second term in the brackets of Eq. (15) to zero coincide with those which were just determined using Eq. (16) where the first term in Eq. (15) is set to zero.

## 7. Limit for the total time-dependent current in an infinite set of adjacent harmonics

The current in each harmonic is zero at the beginning or end of each laser period. However, the total current has a dominant positive peak near the beginning and a dominant negative peak near the end with additional positive and negative peaks having lower magnitude throughout each laser period. Increasing the total number of the harmonics increases the magnitude of the dominant peaks while increasing the rate of the decay of the consecutive extrema when approaching the center of each period. Other phenomena occur near the cutoff that are not considered in this section.

No one has yet measured the time-dependent current at the harmonics, but now we use the identity in Eq. (17) [12] to derive an approximate expression for the current when it is well below the cutoff.

$$\sum_{k=1}^{\infty} \sin(kA) \equiv \frac{\sin(C)}{2[1-\cos(C)]} \text{ for } C > 0 \quad (17)$$

We use Eq. (18) for the current in the frequency domain to include all of the harmonics, but again, we neglect the effects of cutoff that are shown in Eq. (10).

$$I(t) = I_1 \sum_{k=1}^{\infty} \sin(k\omega_0 t) \quad (18)$$

We use Eq. (18) to show that "C" in Eq. (17) is equal to $\omega_0 t$ and then combine Eq. (17) with Eq. (18) to obtain Eq. (19) as an approximation for the total current as a function of time. By inspection, Eq. (19) is singular when $\omega_0 t$ is an even integer and zero when $\omega_0 t$ is an odd integer. Thus, the Fourier series in Eq. (18) is singular at these values but this is because we have assumed an infinite series without including the effects of the cutoff.

$$I(t) = \frac{I_1 \sin(\omega_0 t)}{2[1-\cos(\omega_0 t)]} \quad (19)$$

## 8. Total RMS current in the harmonics within the nanocircuit well below the cutoff

We have made large sets of numerical simulations in which $k_1$ and $k_2$ were specified and Eq. (A1-3) in Appendix 1 was used to obtain sequences of values for the ratio of the normalized time-dependent current, defined as $I(t)/I_1$, at evenly-spaced times covering the full range from $0 < \omega_0 t < 2\pi$. This is one full period at the laser pulse repetition rate $\omega_0$ so these currents would be repeated ad-infinitum with an ideal laser.

To clarify the notation, in Eq. (20) we define the symbol "K" to represent the total number of adjacent harmonics. The sum of the squares for the ratio $I(t)/I_1$ at each specified time was calculated and divided by the number of samples to determine the mean value. Thus, the RHS of Eq. (21) is one-half of the number of harmonics and Eq. (22) gives the normalized RMS value of the total current in the full set of the harmonics from $k_1$ to $k_2$.

$$K \equiv k_2 - k_1 + 1 \tag{20}$$

$$\overline{\left(\frac{I_{kn_1,n_2}}{I_1}\right)^2} = \frac{K}{2} \tag{21}$$

$$\frac{I_{k_1,k_2,RMS}}{I_1} = \sqrt{\frac{K}{2}} \tag{22}$$

Equation (21) shows that the total power delivered to a load with resistance $R_L$ is given by Eq. (23) which is exactly the product of the total number of harmonics in this set multiplied by the power that would be delivered at the first harmonic because $I_1$ is the peak value of the current at that harmonic. However, the transmitting antenna in the nanocircuit has a frequency-dependent impedance consisting of its loss-resistance, radiation-resistance, and reactance so Eq. (23) does not give the full answer for the radiated power. In Section 13 we consider one possibility for this antenna.

$$P_R = K \frac{I_1^2 R_L}{2} \tag{23}$$

## 9. Transmitting antennas for the nanocircuit

We will list several types of the transmitting antennas that may be used in different applications for these nanocircuits. First, we consider the dipole antenna shown in Fig. 2. The radiation resistance $R_R$ of a thin-wire dipole antenna with length L is given by Eq. (24) [13] which we have verified. Here $\eta \approx 376.73$ $\Omega$ is the impedance of free-space, $\gamma \approx 0.57722$ is Euler's constant, $\lambda$ is the wavelength, and $C_i(x)$ and $S_i(x)$ are the cosine and sine integral functions. The load resistance $R_L$ of the antenna, which depends on the resistivity and the dimensions of the dipole has been plotted with the radiation resistance, $R_R$, in the reference by Balanis [13].

$$R_R = \frac{\eta}{2\pi} \left\{ \begin{array}{l} \gamma + \ln\left(\frac{2\pi L}{\lambda}\right) - C_i\left(\frac{2\pi L}{\lambda}\right) + \frac{1}{2}\sin\left(\frac{2\pi L}{\lambda}\right)\left[S_i\left(\frac{4\pi L}{\lambda}\right) - 2S_i\left(\frac{2\pi L}{\lambda}\right)\right] \\ + \frac{1}{2}\cos\left(\frac{2\pi L}{\lambda}\right)\left[\gamma + \ln\left(\frac{\pi L}{\lambda}\right) + C_i\left(\frac{4\pi L}{\lambda}\right) - 2C_i\left(\frac{2\pi L}{\lambda}\right)\right] \end{array} \right\} \tag{24}$$

Equation (24) shows that the radiation resistance and the load resistance are both maximum when the dipole-length is an even-multiple of half a wavelength, and minimum when the dipole-length is an odd-multiple of half a wavelength. However, the ratio of the load resistance to the radiation resistance at their common maxima becomes large at even number of half-wavelengths and near unity at odd multiples of one half-wavelength. Thus, the transmitted power, which is proportional to the ratio of the ratio of $R_R$ to $R_L$, is maximum when the dipole-length is an odd multiple of one-half-wavelength and minimum for even multiples of one-half wavelength. This relationship may be used to divide the harmonics into groups. Others have used long-wire antennas consisting of a gold thin-film line making an ohmic contact on a semiconductor substrate to detect $CO_2$ laser radiation [14]. However, they only used the laser frequency so they did not utilize the broad-band features that are possible with this type of antenna.

Even at the cutoff frequency of 45 THz each monopole in a half-wave dipole must have a length of 1.7 μm so this would not be suitable in a nanocircuit. Thus, we consider the possibility of having dipoles with a much smaller length. Using asymptotic expansions of the Ci and Si functions in Eq. (24) gives Eq. (25) which is generally used by others [15] as the radiation resistance of a short dipole antenna.

$$R_r = \frac{\pi\eta}{6}\left(\frac{L}{\lambda}\right)^2 \quad (25)$$

It should be recognized that the combination of a high-power flux density from the laser and the unusually high signal-to-noise ratio that we have already measured in the harmonics with an STM may enable the detection of the power that is transmitted by antennas even when these antennas are unusually short with a very small radiation resistance. Furthermore, the components of the nanocircuits will have a size that is considerably less than that of the tip and the sample electrodes in an STM which will increase the ratio of the surface area to the mass to provide much greater heat transfer, thus enabling the use of much higher current densities which will increase the radiated power.

A variety of electrically-small antennas have been described [16] which may be considered for use in place of the linear dipole that we showed in Fig. 2. We are particularly interested in the recent simulations of the "frequency-independent" square Archimedean spiral antenna [17] because it may be simpler to fabricate in the nanocircuits. This is one of many possibilities for increasing the usable bandwidth of an antenna by using a two-dimensional design instead of a linear antenna. Simulations of the square Archimedean spiral antenna [17], as well as the recent simulations and measurements that have been made for the traditional circular Archimedean spiral [18], [19], [20], show that these antennas have unusually large usable bandwidths so it is surprising that the Archimedean spiral antenna was not mentioned in the extensive summary of small antennas by Hansen and Collin [16].

The configuration that is shown in Fig. 4 could be used in the nanocircuits as shown in the sketch for Fig. 5. Our calculations suggest that by using many turns in a small area the radiation resistance for this antenna exceeds that of a linear dipole shown in Fig. 2. The resistor at the center of the antenna in Fig. 4 and Fig. 5 corresponds to the resistor that is shown in Fig. 2, but now this resistor may not be required.

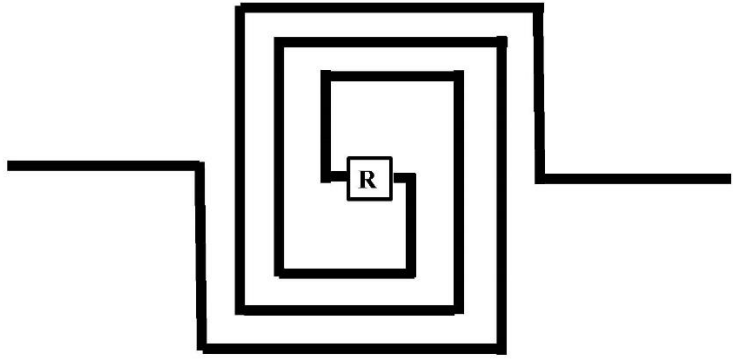

Fig. 4. Archimedean spiral antenna with resistive load at the center.

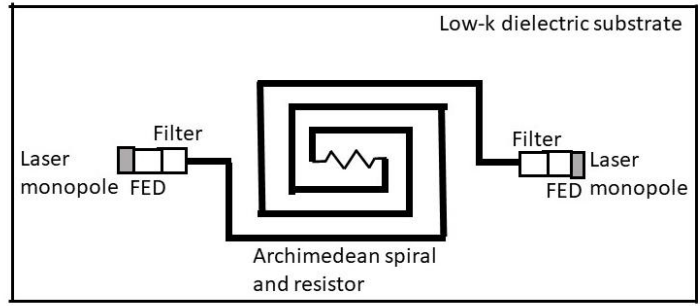

Fig. 5. Sketch of the nanocircuit with an Archimedean spiral antenna

## 10. Presentation of simulations

The simulations for examples 1 through 7 were made by using analytical expressions for the field emission current to follow this current by time-stepping over one or more periods of the mode-locked laser. Example 8 is an analytical solution corresponding to an infinite number of adjacent harmonics. In each of the eight examples the results are presented for one period of the oscillations so the time satisfies $0 \leq \omega_0 t \leq 2\pi$ radians. A discussion of the phenomena in these eight examples follows at the end of their presentation.

Example 1 (Fig. 6) shows the total normalized current $I(t)/I_1$ in the 10 adjacent harmonics where $k_1$ is 1 and $k_2$ is 10 as calculated using Eq. (A1-3) in Appendix 1. There are 10 consecutive pairs or minima and maxima in Fig. 7 which is equal to the number of harmonics and this effect is also seen in Figs. 8 through 11. This is consistent with using Eq. (16) in the body of this paper to predict the number of nulls in a given set of adjacent harmonics.

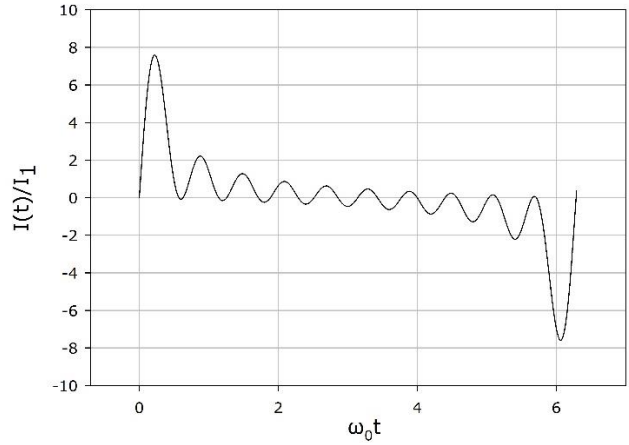
Fig. 6. Normalized total current for 10 adjacent harmonics with indices from 1 through 10.

Example 2 (Fig. 7) shows the total normalized current $I(t)/I_1$ in the 50 adjacent harmonics where $k_1$ is 1 and $k_2$ is 50 as calculated using Eq. (12).

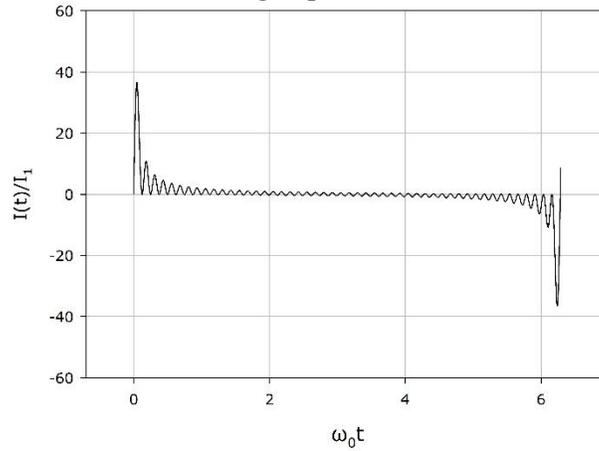
Fig. 7. Normalized total current for 50 adjacent harmonics with indices from 1 through 50.

Example 3 (Fig. 8) shows the total normalized current $I(t)/I_1$ in the 100 adjacent harmonics where $k_1$ is 1 and $k_2$ is 100 as calculated using Eq. (12).

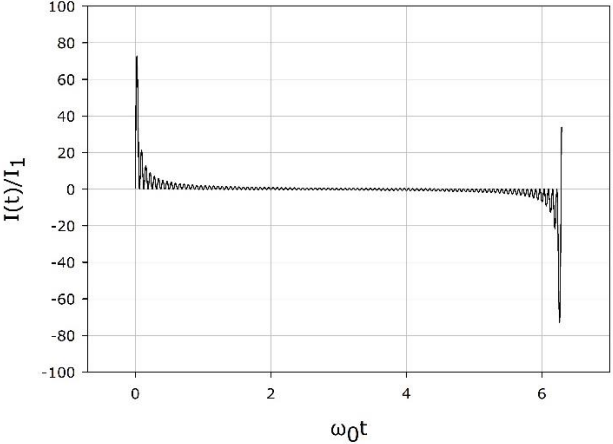

Fig 8. Normalized total current for 100 adjacent harmonics with indices from 1 through 100.

Example 4 (Fig. 9) shows the total normalized current $I(t)/I_1$ in the 500 adjacent harmonics where $k_1$ is 1 and $k_2$ is 500 as calculated using Eq. (12).

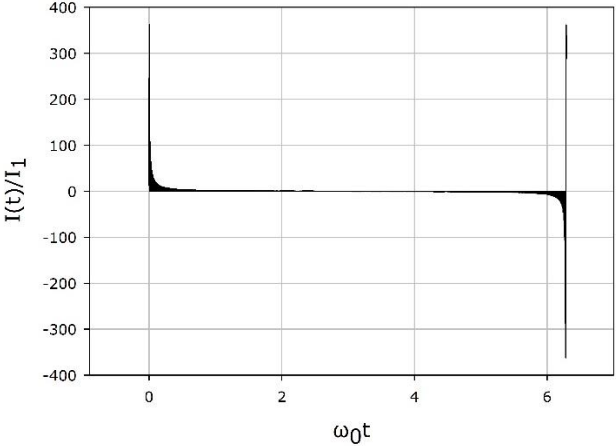

Fig. 9. Normalized total current for 500 adjacent harmonics with indices from 1 through 500.

Example 5 (Fig. 10) shows the total normalized current $I(t)/I_1$ in the 20 adjacent harmonics where $k_1$ is 100 and $k_2$ is 120 as calculated using Eq. (12).

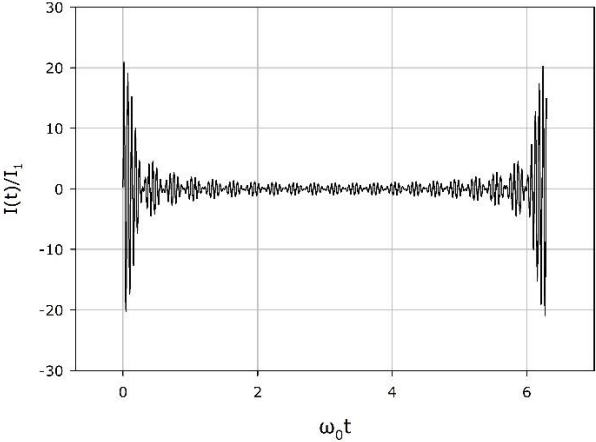

Fig. 10. Normalized total current for 20 adjacent harmonics with indices from 100 through 120.

Example 6 (Fig. 11) shows the total normalized current $I(t)/I_1$ in the 20 adjacent harmonics where $k_1$ is 1000 and $k_2$ is 1020 as calculated using Eq. (12).

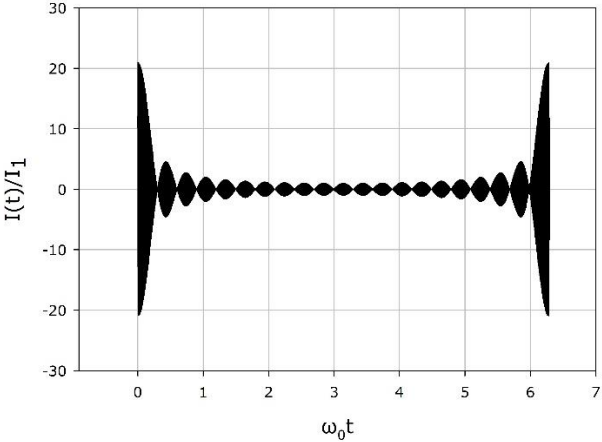

Fig. 11. Normalized total current for 20 adjacent harmonics with indices from 1000 through 1020.

Example 7 (Fig. 12) shows the total normalized current $I(t)/I_1$ in the 200 adjacent harmonics where $k_1$ is 1000 and $k_2$ is 1200 as calculated using Eq. (12).

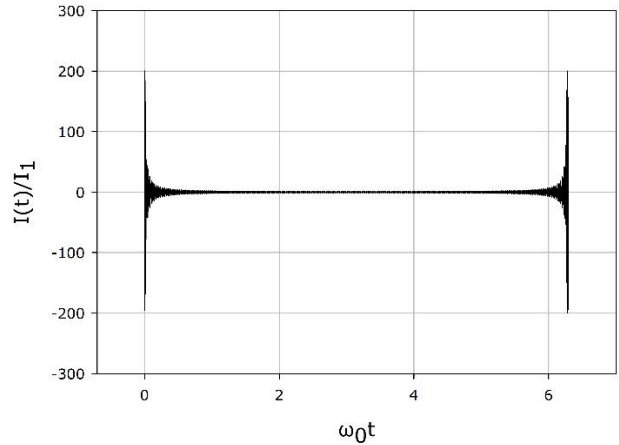

Fig. 12. Normalized total current for 200 adjacent harmonics with indices from 1000 through 1200.

Example 8 (Fig. 13) shows the total normalized current $I(t)/I_1$ determined using Eq. (56). This corresponds to an infinite number of adjacent harmonics with indices from 1 to infinity. However, the effects of the exponential roll-off of the harmonics at the cutoff are not included in this calculation. Thus, there are singularities in place of the sharp rise and fall at the ends of the period.

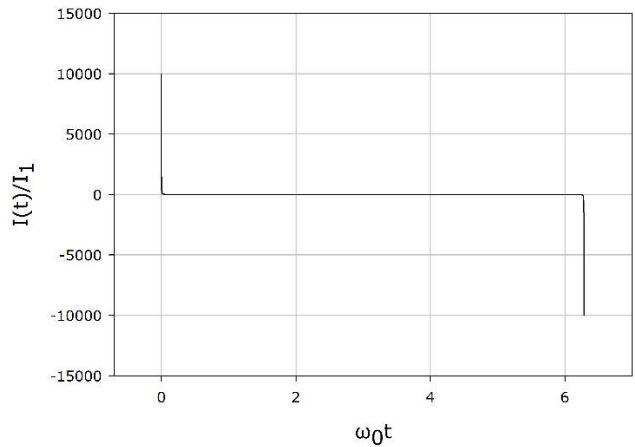

Fig. 13. Normalized total current calculated with Eq. 56 corresponding to an infinite number of adjacent harmonics.

In examples 1 through 4, Figs. 7, 8, 9 and 10, where the adjacent harmonics begin with $k_1$ equal to 1, the total current has a positive maximum near the beginning and a negative extremum near the end of each period. This may be understood because all of the sinusoids have zero amplitude at both ends and they have different frequencies. Thus, increasing the number of harmonics increases the magnitudes of the extrema at the ends as well as the rate of decay toward the center of each period. In example 8, which approximates having an infinite number of adjacent harmonics, there are singularities at both ends of each period.

In examples 5, 6, and 7, Figs. 11, 12, and 13 each have a group of sinusoidal pulses that are caused by $k_1$ being different from 1. In example 5, shown in Fig. 11, $k_1$ is 100 and $k_2$ is 120 so $k_2$ minus $k_1$ is 20 and there are 20 sinusoidal pulses. In example 6, shown in Fig. 12, $k_1$ is 1000 and $k_2$ is 1020 so $k_2$ minus $k_1$ is 20 and there are again 20 sinusoidal pulses. In example 7, shown in Fig. 13, $k_1$ is 1000 and $k_2$ is 1200 so $k_2$ minus $k_1$ is 200 and there are 200 sinusoidal pulses but they are difficult to see in this figure because they are small. The result in each of these three examples may be understood by considering the summation from $k_1$ to $k_2$ in Eq. (49) as the sum from 1 to $k_2$ minus the sum from 1 to $k_1$ as shown in Eq. (60).

$$I(t) = I_1 \sin\left[(k_2+1)\frac{\omega_0 t}{2}\right] \left[\cos\left[(k_2-1)\frac{\omega_0 t}{2}\right] + \sin\left[(k_2-1)\frac{\omega_0 t}{2}\right]\cot\left(\frac{\omega_0 t}{2}\right)\right]$$
$$- I_1 \sin\left[(k_1+1)\frac{\omega_0 t}{2}\right] \left[\cos\left[(k_1-1)\frac{\omega_0 t}{2}\right] + \sin\left[(k_1-1)\frac{\omega_0 t}{2}\right]\cot\left(\frac{\omega_0 t}{2}\right)\right] \quad (60)$$

In examples 1 through 7 Eq. (12) was used to make independent calculations of the total current at a large number of time-steps on consecutive rows of an Excel file over one full cycle of the laser with corresponding to times from zero to $2\pi/\omega_0$. While time-steps of equal size were used in each simulation this is not necessary because each calculation is independent. Furthermore, it would be possible to repeat the calculations at each time-step with different pairs of values for $k_1$ and $k_2$ and then add the results at each time-step to determine the total current for the superposition of several groups of adjacent harmonics. It should be clear that all harmonics are inherently present but specifying values of $k_1$ and $k_2$ is necessary to determine which are passed by the filters as well as the effects of the transmitting antenna.

Table 1 gives the values of $\omega_0 t$ and the normalized current $I/I_1$ that correspond to the first peak in the current for sets of adjacent harmonics calculated using Eq. (49) with specified values of the parameters $k_1$ and $k_2$. The values in this table show that as the number of harmonics is increased the first peak moves toward the left end of the period and the magnitude of this peak becomes greater as may be seen in the Figs. 7 through 13.

Table 1. First peak in the total current with sets of adjacent harmonics.

| $k_1$ | $k_2$ | $\omega_0 t$ | Max $I/I_1$ |
|---|---|---|---|
| 1 | 2 | 0.93593 | 1.7602 |
| 1 | 3 | 0.66729 | 2.4996 |
| 1 | 4 | 0.51861 | 3.2324 |
| 1 | 5 | 0.42416 | 3.9622 |
| 1 | 6 | 0.35883 | 4.6904 |
| 1 | 7 | 0.310943 | 5.4176 |
| 1 | 8 | 0.27434 | 6.1442 |
| 1 | 9 | 0.24544 | 6.8704 |
| 1 | 10 | 0.22206 | 7.5963 |
| 1 | 11 | 0.20274 | 8.3220 |
| 1 | 12 | 0.18652 | 9.0475 |

The creation of intense currents having opposite polarities in the sudden transitions at the start and finish of each laser pulse, is expected to result in partial cancellation of the harmonics which may reduce the output power in the limit as many harmonics are used. This effect may be mitigated in practice by following one or more of the following three procedures: The first method is to reduce the coherence length of the laser as described at the end of Section 9. The second method is to use a moderate number of harmonics chosen for maximum power with stable operation. The third method is to use multiple non-overlapping sets of harmonics. We anticipate that shortening the coherence length of the laser would also increase the power at the harmonics in our measurements with a scanning tunneling microscope but we have not had the opportunity to test this hypothesis. For example, this method could be applied by using harmonics near the different maxima of the radiation resistance for a long-wire antenna that were described in Section 13.

## 11. Suggestions for optimizing the nanocircuits in microwave and terahertz applications

It is possible to transmit one or more harmonics and the radiated power is increased by raising this number. At microwave frequencies filters can permit the transmission of only a single harmonic or several chosen harmonics. However, groups of adjacent harmonics must be used at terahertz frequencies because of fundamental limitations in the sharpness of the cutoffs for filters. The long-wire antennas described in Section 13 may be used to transmit groups of 1 or more microwave harmonics at frequencies where Eq. (58) shows that the length of the dipole is near an even multiple of one-half of the wavelength. Long-wire antennas could also be used to transmit one or more selected groups of terahertz harmonics at frequencies where the dipole-length is near an even multiple of one-half of the wavelength.

It is possible to use different types of sensors as parts in the two filters of the nanocircuit. These filters could also modulate the harmonics in order to transmit data such as local values for the temperature, ionizing radiation, presence of specific chemicals, or other phenomena, so that one, or even many, nanocircuits could provide such data at high-speed with extremely fine spatial resolution.

## 12. Summary and conclusions

Our measurements in laser-assisted scanning tunneling microscopy show that focusing a mode-locked laser on the tip-sample tunneling junction generates a set of harmonics at integer multiples of the laser pulse repetition frequency. Analysis of the data from these measurements shows that the measured roll-off in the power of these harmonics as the square of the frequency is caused by the circuit that is used in these measurements, and further analysis suggests that these harmonics continue until reaching a cutoff at approximately 45 THz for a total of approximately $6 \times 10^5$ harmonics.

We have designed a nanocircuit which would mitigate the measured roll-off so that the current may be approximately the same up to the cutoff and have already made progress in developing the field emission diodes which we consider to be the most challenging part of this project. As was already stated in the abstract, in this unique approach, when transmitting over a chosen bandwidth $\Delta f$ that is proportional to $f_0$, the frequency at the center of this bandwidth, the output power is proportional to the square of the center frequency because of the fixed spacing of the harmonics.

**Acknowledgments**

We are grateful to the Center for Integrated Nanotechnologies (CINT) at Los Alamos National Laboratory (LANL) for making it possible for Dr. Hagmann to study laser-assisted tunneling in laser-assisted scanning tunneling microscopy in semi-annual two-week visits from 2008 to 2017. This enabled experimental verification and characterization of the MFC generated using a mode-locked laser with an STM.

We acknowledge support by the NSF STTR grant 0712564 "Fabrication and testing of laser-assisted terahertz field emitters". This enabled the author as the PI to work with Professors Ajay Nahata and Mark Miller at the University of Utah [21] to develop field emission diodes with stable operation in air at atmospheric pressure [22] that we will develop further for the nanoscale circuit. The author plans to continue this project in further research and development in his recent appointment as a Research Professor in Electrical and Computer Engineering at the University of Utah. Isaac Martin, a contractor to NewPath Research L.L.C., verified and clarified the interpretation of Eq. (24) that was presented earlier by Balanis [13] and prepared the graphs for this publication.

**Data availability statement**

The author has chosen to make data that support the findings of this study available upon reasonable request.
**References**

1. M.J. Hagmann, "Simulations of the interaction of tunneling electrons with optical fields in laser-illuminated field emission", J. Vac. Sci. Technol. B 13 (1995) 1348-1352.
2. M. J. Hagmann, "Mechanism for resonance in the interaction of tunneling particles with modulation quanta", J. Appl. Phys.78 (1995) 25-29.
3. M.J. Hagmann and M. Brugat, "Modulation of the current in a field emitter caused by a continuous wave or pulsed laser: Simulations and experimental results", J. Vac. Sci. Technol. B 15 (1997) 405-409.
4. M.J Hagmann, "Simulations of the generation of broadband signals from DC to 100 THz by photomixing in laser-assisted field emission", Ultramicroscopy 73 (1998) 89-97.
5. M.J. Hagmann, A.J. Taylor and D.A Yarotski, "Observation of $200^{th}$ harmonic with fractional linewidth of $10^{-10}$ in a microwave frequency comb generated in a tunneling Junction", Appl. Phys. Lett. 101 (2012) 241102.
6. M.J. Hagmann, F.S. Stenger and D.A Yarotski, "Linewidth of the harmonics in a microwave frequency comb generated by focusing a mode-locked ultrafast laser on a tunneling junction", J. Appl. Phys. 114 (2013) 223107.
7. M.J. Hagmann, D.G. Coombs and D.A. Yarotski, "Periodically pulsed laser-assisted tunneling may generate terahertz radiation", J. Vac. Sci. Technol. B 35 (2017) 03D109.
8. M.J. Hagmann, A. Efimov, A.J. Taylor, and D.A. Yarotski, "Microwave frequency-comb generation in a tunneling junction by intermode mixing of ultrafast laser pulses", Appl. Phys. Lett. 99, 011112 (2011).
9. M.J. Hagmann, P. Andrei, S. Pandey and A. Nahata, "Possible applications of scanning frequency comb microscopy for carrier profiling in semiconductors", J. Vac. Sci. Technol. B, 33 (2015) 02B109 (6 pp).
18

**Appendix 1. Total current in a set of adjacent harmonics well below the cutoff**

As already noted in the main body of this paper, at the higher harmonics, the exponential in Eq. (10) causes a cutoff where the current decays to zero. However, in this appendix we make the approximation of setting the exponential to unity to obtain a closed-form expression for the current in a set of harmonics that are well below this cutoff. Thus, we consider Eq. (A1-1).

$$I(t) = I_1 \sum_{k=k_1}^{k_2} \sin(k\omega_0 t) \qquad (A1-1)$$

We have derived the identity in Eq. (A1-2) by using Wolfram Alpha, verified it, and then used it with Eq. (A1-1) to obtain Eq. (A1-3) as an approximation for the total current as a function of time for all of the harmonics in the interval $k_1 \leq k \leq k_2$. The numerical testing included several examples where the terms for each value of k were calculated individually at different time-steps using the argument of Eq. (A1-1) and then summed. These sums, including each time-step, were then found to be equal to the closed-form solution in Eq. (A1-3).

$$\sum_{k=k_1}^{k_2} \sin(kx) \equiv \frac{1}{2}\left[\begin{array}{l}\sin(k_1 x)+\cot\left(\frac{x}{2}\right)\cos(k_1 x) \\ +\sin(k_2 x)-\cot\left(\frac{x}{2}\right)\cos(k_2 x)\end{array}\right] \quad (A1-2)$$

$$I(t) = \frac{I_1}{2}\left[\begin{array}{l}\sin(k_1\omega_0 t)+\cot\left(\frac{\omega_0 t}{2}\right)\cos(k_1\omega_0 t) \\ +\sin(k_2\omega_0 t)-\cot\left(\frac{\omega_0 t}{2}\right)\cos(k_2\omega_0 t)\end{array}\right] \quad (A1-3)$$

**Appendix 2. Derivation of the current in a set of adjacent harmonics near the cutoff**

Let $\|I_k\|$ be defined as the peak value for the current at the kth harmonic. Thus, because the spacing between adjacent harmonics is small near the cutoff, the value for the peak of the k+1 harmonic is related to that for the k-th harmonic by the approximation in Eq. (A2-1):

$$\frac{\|I_{k+1}\|}{\|I_k\|} = \frac{e^{-A(k+1)^2}}{e^{-Ak^2}} \approx e^{-A(2k+1)} \quad (A2-1)$$

The Ti:Sapphire laser used in our measurements at Los Alamos has a pulse duration $\tau$ of 6 fs, and a pulse repetition rate of 74.25 MHz so the time between consecutive pulses T is 13.5 ns [A2-1]. Thus, from Eq. 8 in the main body of this paper, the dimensionless parameter A is 9.7 x10$^{-13}$. We label the value of k at which the peak power has decayed to one-half of that at the fundamental as $k_{3dB}$ to obtain Eq. (A2-2), and write Eq. (A2-3) for the current at the fundamental where k is equal to 1.

$$\|I_{k_{3dB}}\| = I_1 e^{-A k_{3dB}^2} \quad (A2-2)$$

$$\|I_1\| = I_1 e^{-A} \quad (A2-3)$$

Equation (A2-4) is obtained by combining Eqs. (A2-2) and (A2-3), so the ratio of these two values of the power is given by Eq. (A2-5). To determine the value of k at the 3-dB point this ratio is set to one-half to obtain Eq. (A2-6) which is rearranged to give Eq. (A2-7), Eq. (A2-8), and thus Eq. (A2-9) is the value for the index k at the 3-dB point.

$$\frac{\|I_{k_{3dB}}\|}{\|I_1\|} = e^{A - A k_{3dB}^2} \quad (A2-4)$$

$$\frac{P_{k_{3dB}}}{P_1} = e^{2A - 2A k_{3dB}^2} \quad (A2-5)$$

$$2e^{2A - 2A k_{3dB}^2} = 1 \quad (A2-6)$$

$$\ln(2) = 2A k_{3dB}^2 - 2A \quad (A2-7)$$

$$\frac{\ln(2)}{2A} + 1 = k_{3dB}^2 \quad (A2-8)$$

$$k_{3dB} = \sqrt{1 + \frac{\ln(2)}{2A}} \quad (A2-9)$$

Earlier in this section we noted that in our measurements of the MFC with an STM the parameter A was approximately 9.7 x$10^{-13}$. Thus, Eq. (A2-9) shows that under these conditions $k_{3dB}$ would be 6.0 x$10^5$ so the cutoff would be near 45 THz.

Next, we consider how to determine the total current as a function of time for a small range of harmonics with $k_1 \le k \le k_2$ where $k_1$ and $k_2$ are large integers. When the range, defined as $k_2$ minus $k_1$, is much smaller than either $k_1$ or $k_2$ there will be little variation in the value of the exponential in Eq. (10) in the main body of the paper. Thus, we approximate Eq. (10) with Eq. (A2-10) where $\bar{k}$ is defined as the mean value for $k_1$ and $k_2$.

Then the summation in Eq. (10) of the body for this paper is used with Eq. (22) simplifies to obtain Eq. (A2-10) as an approximation for the total current in the harmonics from $k_1$ to $k_2$. This is simplified by using Eq. (A1-2) to obtain Eq. (A2-11). However, we anticipate that actual measurements of the spectrum of the harmonics that are close to the cutoff may be sensitive to the stability of the mode-locked laser.

$$I(t) = I_1 e^{-A\bar{k}^2} \sum_{k=k_1}^{k=k_2} \sin(k\omega_0 t) \qquad (A2-10)$$

$$I(t) = \frac{I_1}{2} e^{-A\bar{k}^2} \begin{bmatrix} \sin(k_1\omega_0 t) + \cot\left(\frac{\omega_0 t}{2}\right)\cos(k_1\omega_0 t) \\ + \sin(k_2\omega_0 t) - \cot\left(\frac{\omega_0 t}{2}\right)\cos(k_2\omega_0 t) \end{bmatrix} \qquad (A2-11)$$

**Appendix 3. Derivation of the current in any set of adjacent harmonics**

The Taylor series expansion for the exponential in Eq. (10) in the body of this paper is given by Eq. (A3-1).

$$e^{-Ak^2} = \sum_{n=0}^{\infty} \frac{(-Ak^2)^n}{n!} \qquad (A3-1)$$

Combining Eqs. (10) and (A3-1) gives Eq. (A3-2) for the total current where x is $\omega_0 t$.

$$I(t) = I_1 \sum_{k=k_1}^{k=k_2} \left[ \sin(kx) \sum_{n=0}^{\infty} \frac{(-Ak^2)^n}{n!} \right] \qquad (A3-2)$$

Writing the first few terms of the summation in Eq. (A3-1) gives Eq. (A3-3) for the total current.

$$I(t) = I_1 \sum_{k=k_1}^{k=k_2} \left[ \sin(kx) \left\{ 1 - \frac{Ak^2}{1!} + \frac{A^2 k^4}{2!} - \frac{A^3 k^6}{3!} + \frac{A^4 k^8}{4!} - \frac{A^5 k^{10}}{5!} + \cdots \right\} \right] \qquad (A3-3)$$

Next, the first six terms of the inner summation in Eq. (A3-3) are separated to give 6 parts where the first one, denoted by $T_1$, is Eq. (A3-4) which was derived earlier in the main body of this paper as Eq. (11). The definitions of the other 5 other terms in Eq. (A3-3), and closed-form expressions from Wolfram Alpha which we have confirmed, are given in Eqs. (A3-5) through (A3-9).

$$T_1 \equiv \sum_{k=k_1}^{k_2} \sin(kx) = \frac{1}{2}\left[\sin(k_1 x) + \cot\left(\frac{x}{2}\right)\cos(k_1 x) + \sin(k_2 x) - \cot\left(\frac{x}{2}\right)\cos(k_2 x)\right] \quad (A3-4)$$

$$T_2 \equiv -\frac{A}{1!}\sum_{k=k_1}^{k=k_2} k^2 \sin(kx) = -\frac{A}{1}x^2 \csc\left(\frac{x}{2}\right)\sin\left[\frac{1}{2}(1-k_1+k_2)x\right]\sin\left[\frac{1}{2}(n_1+n_2)x\right] \quad (A3-5)$$

$$T_3 \equiv \frac{A^2}{2!}\sum_{k=k_1}^{k=k_2} k^4 \sin(kx) = \frac{A^2}{2!}x^4 \csc\left(\frac{x}{2}\right)\sin\left[\frac{1}{2}(1-k_1+k_2)x\right]\sin\left[\frac{1}{2}(k_1+k_2)x\right] \quad (A3-6)$$

$$T_4 \equiv -\frac{A^3}{3!}\sum_{k=k_1}^{k=k_2} k^6 \sin(kx) = -\frac{A^3}{3!}x^6 \csc\left(\frac{x}{2}\right)\sin\left[\frac{1}{2}(1-k_1+k_2)x\right]\sin\left[\frac{1}{2}(k_1+k_2)x\right] \quad (A3-7)$$

$$T_5 \equiv \frac{A^4}{4!}\sum_{k=k_1}^{k=k_2} k^8 \sin(kx) = \frac{A^4}{4!}x^8 \csc\left(\frac{x}{2}\right)\sin\left[\frac{1}{2}(1-k_1+k_2)x\right]\sin\left[\frac{1}{2}(k_1+k_2)x\right] \quad (A3-8)$$

$$T_6 \equiv -\frac{A^5}{5!}\sum_{k=k_1}^{k=k_2} k^{10} \sin(kx) = -\frac{A^5}{5!}x^{10} \csc\left(\frac{x}{2}\right)\sin\left[\frac{1}{2}(1-k_1+k_2)x\right]\sin\left[\frac{1}{2}(k_1+k_2)x\right] \quad (A3-9)$$

Setting x to $\omega_0 t$ and simplifying Eqs. (A3-4) to (A3-9) gives Eqs. (A3-10) through (A3-15).

$$T_1 = \frac{1}{2}\left[\sin(k_1\omega_0 t) + \cot\left(\frac{\omega_0 t}{2}\right)\cos(k_1\omega_0 t) + \sin(k_2\omega_0 t) - \cot\left(\frac{\omega_0 t}{2}\right)\cos(k_2\omega_0 t)\right] \quad (A3-10)$$

$$T_2 = -\frac{A}{1}(\omega_0 t)^2 \csc\left(\frac{\omega_0 t}{2}\right)\sin\left[\frac{1}{2}(1-k_1+k_2)\omega_0 t\right]\sin\left[\frac{1}{2}(k_1+k_2)\omega_0 t\right] \quad (A3-11)$$

$$T_3 = \frac{A^2}{2!}(\omega_0 t)^4 \csc\left(\frac{\omega_0 t}{2}\right)\sin\left[\frac{1}{2}(1-k_1+k_2)\omega_0 t\right]\sin\left[\frac{1}{2}(k_1+k_2)\omega_0 t\right] \quad (A3-12)$$

$$T_4 = -\frac{A^3}{3!}(\omega_0 t)^6 \csc\left(\frac{\omega_0 t}{2}\right)\sin\left[\frac{1}{2}(1-k_1+k_2)\omega_0 t\right]\sin\left[\frac{1}{2}(k_1+k_2)\omega_0 t\right] \quad (A3-13)$$

$$T_5 = \frac{A^4}{4!}(\omega_0 t)^8 \csc\left(\frac{\omega_0 t}{2}\right)\sin\left[\frac{1}{2}(1-k_1+k_2)\omega_0 t\right]\sin\left[\frac{1}{2}(k_1+k_2)\omega_0 t\right] \quad (A3-14)$$

$$T_6 = -\frac{A^5}{5!}(\omega_0 t)^{10} \csc\left(\frac{\omega_0 t}{2}\right)\sin\left[\frac{1}{2}(1-k_1+k_2)\omega_0 t\right]\sin\left[\frac{1}{2}(k_1+k_2)\omega_0 t\right] \quad (A3-15)$$

Thus, the total current as a function of time is obtained by multiplying $I_1$ by the sum of the RHS in Eqs. (A3-10) through (A3-15) to obtain Eq. (A3-16), recognizing that Eq. (A3-16) must be continued as shown by the ellipsis, where this is factored to obtain Eq. (A3-17).

$$I(t) = \frac{I_1}{2}\left[\sin(k_1\omega_0 t) + \cot\left(\frac{\omega_0 t}{2}\right)\cos(k_1\omega_0 t) + \sin(k_2\omega_0 t) - \cot\left(\frac{\omega_0 t}{2}\right)\cos(k_2\omega_0 t)\right]$$

$$-\frac{AI_1}{1!}(\omega_0 t)^2 \csc\left(\frac{\omega_0 t}{2}\right)\sin\left[\frac{1}{2}(1-k_1+k_2)\omega_0 t\right]\sin\left[\frac{1}{2}(k_1+k_2)\omega_0 t\right]$$

$$+\frac{A^2 I_1}{2!}(\omega_0 t)^4 \csc\left(\frac{\omega_0 t}{2}\right)\sin\left[\frac{1}{2}(1-k_1+k_2)\omega_0 t\right]\sin\left[\frac{1}{2}(k_1+k_2)\omega_0 t\right]$$

$$-\frac{A^3 I_1}{3!}(\omega_0 t)^6 \csc\left(\frac{\omega_0 t}{2}\right)\sin\left[\frac{1}{2}(1-k_1+k_2)\omega_0 t\right]\sin\left[\frac{1}{2}(k_1+k_2)\omega_0 t\right]$$

$$+\frac{A^4 I_1}{4!}(\omega_0 t)^8 \csc\left(\frac{\omega_0 t}{2}\right)\sin\left[\frac{1}{2}(1-k_1+k_2)\omega_0 t\right]\sin\left[\frac{1}{2}(k_1+k_2)\omega_0 t\right]$$

$$-\frac{A^5 I_1}{5!}(\omega_0 t)^{10} \csc\left(\frac{\omega_0 t}{2}\right)\sin\left[\frac{1}{2}(1-k_1+k_2)\omega_0 t\right]\sin\left[\frac{1}{2}(k_1+k_2)\omega_0 t\right] + \cdots \quad (A3-16)$$

$$I(t) = \frac{I_1}{2}\left[\sin(k_1\omega_0 t) + \cot\left(\frac{\omega_0 t}{2}\right)\cos(k_1\omega_0 t) + \sin(k_2\omega_0 t) - \cot\left(\frac{\omega_0 t}{2}\right)\cos(k_2\omega_0 t)\right]$$

$$-I_1\left[\frac{A}{1!}(\omega_0 t)^2 - \frac{A^2}{2!}(\omega_0 t)^4 + \frac{A^3}{3!}(\omega_0 t)^6 - \frac{A^4}{4!}(\omega_0 t)^8 + \frac{A^5}{5!}(\omega_0 t)^{10} - \cdots\right]$$

$$\csc\left(\frac{\omega_0 t}{2}\right)\sin\left[\frac{1}{2}(1-k_1+k_2)\omega_0 t\right]\sin\left[\frac{1}{2}(k_1+k_2)\omega_0 t\right] \quad (A3-17)$$

We use the Taylor series in Eq. (A3-18), rearranged as Eq. (A3-19), with the argument x changed to obtain Eq. (A3-20). Then this result is used to change the second line of Eq. (A3-17) in order to obtain Eq. (A3-21).

$$e^{-x} = 1 - \frac{x}{1!} + \frac{x^2}{2!} - \frac{x^3}{3!} + \frac{x^4}{4!} - \frac{x^5}{5!} + \cdots \quad (A3-18)$$

$$1 - e^{-x} = \frac{x}{1!} - \frac{x^2}{2!} + \frac{x^3}{3!} - \frac{x^4}{4!} + \frac{x^5}{5!} + \cdots \quad (A3-19)$$

$$1 - e^{-A(\omega_0 t)^2} \equiv \frac{A(\omega_0 t)^2}{1!} - \frac{A^2(\omega_0 t)^4}{2!} + \frac{A^3(\omega_0 t)^6}{3!} - \frac{A^4(\omega_0 t)^8}{4!} + \frac{A^5(\omega_0 t)^{10}}{5!} - \cdots \quad (A3-20)$$

$$I(t) = \frac{I_1}{2}\left[\sin(k_1\omega_0 t) + \cot\left(\frac{\omega_0 t}{2}\right)\cos(k_1\omega_0 t) + \sin(k_2\omega_0 t) - \cot\left(\frac{\omega_0 t}{2}\right)\cos(k_2\omega_0 t)\right]$$

$$-I_1\left[1 - e^{-A(\omega_0 t)^2}\right]\csc\left(\frac{\omega_0 t}{2}\right)\sin\left[\frac{1}{2}(1-k_1+k_2)\omega_0 t\right]\sin\left[\frac{1}{2}(k_1+k_2)\omega_0 t\right] \quad (A3-21)$$

Equation (A3-21) gives the total current as a function of time for a set of adjacent harmonics in any interval for which $k_1 \leq k \leq k_2$ with an ideal mode-locked laser that is started at time t equal to zero and continues with infinite coherence length to infinity. Note that the second term differs in that decays to zero leaving only the first term which repeats the interval for $\omega_0 t$ from 0 to $2\pi$.

However, for an actual mode-locked laser which has a finite coherence length the process for generating the harmonics restarts at the end of each burst of pulses from the laser. Thus, both terms in Eq. (A3-21) must be used repeatedly to model each restart of lasing. In the examples that are shown in Section 14 it will be seen that the total current with a set of adjacent harmonics is maximum at the start and end of each cycle of the laser. Thus, shortening the coherence length of a mode-locked laser may actually increase the output power at the harmonics by increasing the relative contribution from that near the beginning and the end of each cycle.

## Appendix4: Second possibility for zeros in the total current

In the main body of this paper, we derived Eq. (15) for the total current in a set of adjacent harmonics with indices from $k_1$ to $k_2$ and Eq. (16) for the set of times at which the first of the two terms within brackets in Eq. (15) is zero. Now we determine if there are other values of $\omega_0 t$ for which the second term in Eq. (15), which is shown as Eq. (A4-1), is zero.

$$\text{Term2} \equiv \cos\left[(k_2 - k_1)\frac{\omega_0 t}{2}\right] + \sin\left[(k_2 - k_1)\frac{\omega_0 t}{2}\right] \cot\left(\frac{\omega_0 t}{2}\right) \quad (A4-1)$$

The definition of K in Eq. (50) is used with Eq. (A4-1) to give Eq. (A4-2).

$$\text{Term2} = \cos\left[(K-1)\frac{\omega_0 t}{2}\right] + \sin\left[(K-1)\frac{\omega_0 t}{2}\right] \cot\left(\frac{\omega_0 t}{2}\right) \quad (A4-2)$$

At the start and end of each cycle of the laser, where $\omega_0 t$ is an integral multiple of $2\pi$, the second part of Eq. (A4-2) is zero because the sine function is zero. However, the first term is plus or minus 1. Therefore, the points at the beginning and end of each cycle are excluded from the search for possible roots from the second term. Thus, we divide by the sine term in Eq. (A4-2) to obtain Eq. (A4-3) as the condition for which the second term of Eq. (49) could possibly be zero.

$$\cot\left[(K-1)\frac{\omega_0 t}{2}\right] = -\cot\left(\frac{\omega_0 t}{2}\right) \quad (A4-3)$$

When x, as defined in Eq. (A4-4), is substituted into Eq. (A4-3) this gives Eq. (A4-5) where N must be an integer of 1 or greater so we search for values of x that satisfy Eq. (A4-5) over the interval $0 \le \omega_0 t \le 2\pi$ which corresponds to one period of the cotangent function from $0 \le x \le \pi$.

$$x \equiv \frac{\omega_0 t}{2} \quad (A4-4)$$

$$\cot\left[(K-1)x\right] + \cot(x) = 0 \quad (A4-5)$$

The following solutions, derived using Wolfram Alpha and then verified numerically by substitution using bisection for high-accuracy. They correspond to the set of values for x between 0 and $\pi$ that satisfy Eq. (A4-5) for each value of K. When K = 1, $x_1 = \pi/3$ and $x_2 = 2\pi/3$. When K = 2, $x_1 = \pi/4$, $x_2 = 2\pi/4$, and $x_3 = 3\pi/4$. When K = 3, $x_1 = 2\pi/10$, $x_2 = 4\pi/10$, $x_3 = 6\pi/10$, and $x_4 = 8\pi/10$. Table A4-1 shows the corresponding normalized times, t/T as fractions, in one period of the mode-locked laser, at which the total current of the harmonics is zero.

Table A4-1. Normalized times when the total current is zero
in one period for the laser as calculated with Eq. (A4-5).

| K | Normalized times for zero current in one period of the laser, t/T |     |     |     |     |     |
|---|-----|-----|-----|-----|-----|-----|
| 1 | 1/3 | 2/3 |     |     |     |     |
| 2 | 1/4 | 2/4 | 3/4 |     |     |     |
| 3 | 1/5 | 2/5 | 3/5 | 4/5 |     |     |
| 4 | 1/6 | 2/6 | 3/6 | 4/6 | 5/6 |     |
| 5 | 1/7 | 2/7 | 3/7 | 4/7 | 5/7 | 6/7 |

The values in Table A4-1, which were determined by setting the second bracketed term in Eq. (15) to zero, are identical to those we derived earlier in Section 9 by setting the first bracketed term in Eq. (15) to zero. Thus, this appendix shows that no additional zeros for the current are caused by the second term.